\documentclass[aps,twocolumn,floats,showpacs]{revtex4}
\usepackage{epsfig}
\usepackage{graphicx}
\usepackage{amsmath}

\newcommand{\beq}{\begin{equation}}
\newcommand{\eeq}{\end{equation}}
\newcommand{\bea}{\begin{eqnarray}}
\newcommand{\eea}{\end{eqnarray}}

\begin{document}

\title{
Nonlinear quantum heat transfer in hybrid structures: Sufficient conditions for thermal rectification}
\author{Lian-Ao Wu$^{1,2}$, Claire X. Yu$^{2}$, Dvira Segal$^{2}$}

\affiliation{$^1$Department of Theoretical Physics and History of
Science, The Basque Country University (EHU/UPV), PO Box 644, 48080
Bilao, Spain}
\affiliation{$^2$Chemical Physics Group, Department of Chemistry and Center for Quantum
Information and Quantum Control, University of Toronto, 80 St. George
street, Toronto, Ontario, M5S 3H6, Canada}
\date{\today}

\begin{abstract}

We present a unified description of heat flow in two-terminal hybrid quantum systems. 
Using simple models, we analytically study nonlinear 
aspects of heat transfer between various reservoirs: metals, solids, and spin
baths, mediated by the excitation/relaxation of a central (subsystem) mode. 
We demonstrate rich nonlinear current-temperature characteristics,
originating from  either the molecular anharmonicity, or the reservoirs 
(complex) energy spectra.
In particular, we establish sufficient conditions
for thermal rectification in two-terminal junctions. We 
identify two classes of rectifiers. In type-A rectifiers the density of states of
the reservoirs are dissimilar. In type-B rectifiers the
baths are identical, but include particles whose statistics differ
from that of the subsystem, to which they asymmetrically couple.
Nonlinear heat flow, and specifically thermal rectification, are thus ubiquitous
effects that could be observed in a variety of systems, phononic, electronic, and photonic.

\end{abstract}

\maketitle

\section{Introduction}

Understanding heat transfer in two-terminal hybrid structures is of
fundamental and practical importance, for controlling transport at
the nanoscale, and for realizing functional devices
\cite{Cahill, Majumdar-rev}. Among the systems that fall into this category
are metal-molecule-metal junctions, the basic component of molecular
electronic devices. Here the excessive energy generated at the
molecular core should be effectively removed for realizing a stable
device, as demonstrated theoretically \cite{heating} 
and experimentally \cite{Tao,C60}.
Another composite structure of fundamental interest is a
dielectric-molecule-dielectric system, where vibrational energy
flowing between the components activates reactivity and controls
dynamics \cite{Strub, Hamm, Dlott}. Phononic junctions are also
captivating for understanding the validity of the Fourier's law of
thermal conduction at the nanoscale \cite{Fourier,ZettlF}.
Single-mode {\it radiative} heat conduction between ohmic metals was
recently detected, manifesting that photon-mediated thermal
conductance is quantized \cite{Pekola}. Other hybrid systems with
interesting thermal properties are electronic spin-nuclear spin
interfaces \cite{Lukin}, metal-molecule-dielectric contacts
\cite{Persson} and metal-superconductor junctions \cite{Rev}.


From the theoretical point of view, the basic challenge here is
to understand the role of nonlinear interactions in determining transport mechanisms,
and specifically, in inducing normal
(Fourier) heat conduction, either within classical laws
\cite{Fourier, Lepri, Dhar} or based on quantum mechanical principles
\cite{ProsenQ,Michel05, Michel06, FourierWS, Tu, Rossini}. 
Treatments employed vary. Ballistic heat transfer 
in phononic systems was explored using the Landauer-scattering formalism \cite{Kirczenow} and within the
generalized Langevin equation \cite{SegalHanggi, SC, Dhar}.
For interacting systems there are only few analytical results, and the techniques employed
include the Kinetic-Boltzmann theory \cite{Spohn}, mode coupling theory \cite{MC1,MC2}, 
the non-equilibrium Green's function method \cite{Green,Wang1,Tu},
classical \cite{Lepri} and mixed classical-quantum \cite{Wang} molecular dynamics simulations, 
and exact quantum simulations on simplified models \cite{Hartree}.


The master equation formalism \cite{Breuer} is another useful tool
for studying quantum heat transfer across nanojunctions \cite{MS1,MS2,MS3}. 
We have recently adopted this method
in the weak system-bath coupling limit, and explored various aspects of heat flow
in simplified toy models \cite{Rectif,  NDR, FourierWS, radiation}.
Basically, we ask ourselves the following question: 
What is the connection between the microscopic Hamiltonian and the
heat current across the system? More specifically,
how can we control the onset of nonlinear current-temperature
behavior \cite{Rectif,NDR}, and what conditions should
the Hamiltonian satisfy for the system to manifest normal 
conduction \cite{FourierWS}?
While our treatment has been typically limited
to a minimal subsystem, the formalism has two main advantages: (i) Both harmonic and anharmonic
systems can be treated within the same footings. 
(ii) Analytical results obtained can pinpoint on the underlying dynamics, 
typically obscured in numerical simulations.

In this work we extend our recent letters
\cite{radiation, Sufficient}, and develop a {\it unified} description of
temperature-driven quantum heat transfer at the nanoscale. We focus
on two-terminal devices, including a central quantized unit
(subsystem) and two  bulk objects (referred to as
terminals/contacts/reservoirs/baths), whose temperatures are externally
controlled. Our description can be applied on several systems,
including electron-hole pair excitations, 
phonons, photons or spins. However, our formalism
does not describe heat flow due to the transport of hot charge
carriers \cite{heating}.

\begin{figure}[htbp]
\hspace{2mm}
{\hbox{\epsfxsize=90mm \epsffile{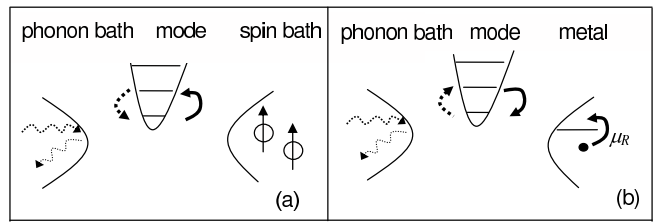}}}
\caption{Examples of two hybrid systems treated in this work.
(a) Single mode heat transfer between a solid and a spin bath.
The central unit can represent either a vibrational  or a radiation  mode.
(b) Phonon to exciton energy exchange. }
\label{Fig0}
\end{figure}

We analytically explore current-temperature 
characteristics in various systems, and seek to connect
the nonlinear behavior to the microscopic parameters. 
We also investigate the temperature dependence of the thermal conductance,
and observe rich behavior, depending of the details of the model.
In particular, we discuss a simple nonlinear effect, 
thermal rectification, an asymmetry of the
heat current for forward and reversed temperature gradients. This
phenomenon has recently attracted considerable theoretical
\cite{Rectif, Terraneo, Casati, Zhang, Prosen, Rectif2D, Rectif3D, RectifMass,Zeng,Lemos,Tomi} 
and experimental \cite{RectifE,QDrectif}
attention. While most theoretical studies have analyzed this
phenomenon in {\it phononic systems} using classical molecular
dynamics simulations, our prototype model can describe thermal
rectification of several subsystems (vibrational or
radiation modes) and reservoirs (spin, metal, dielectrics)  at the same footing,
see Fig. \ref{Fig0}.
Moreover, using our simple model we establish {\it sufficient}
conditions for rectification \cite{Sufficient}. 
We identify two classes of rectifiers: 
(i) "Type A", where the reservoirs are dissimilar i.e. of different
density of states (DOS). (ii) "Type B", where the baths are identical, but their
statistics differ from that of the subsystem, combined with unequal
coupling strengths at the two ends, as explained below. 

The content of the paper is as follows. In Section II we present
our model. In Section III we derive equations of motion for the
subsystem population using the quantum master equation approach.
We consider several models for the subsystem and for the contacts,
and derive analytical expressions for the heat current through various conducting junctions.
In Section IV we trace these junctions' (nonlinear) current-temperature 
characteristics to the Hamiltonian parameters.
In particular, Section V is focused on a specific nonlinear effect, thermal rectification. We
analytically identify two types of rectifiers, 
and demonstrate with numerical simulations the tunability of the effect.
Conclusions follow in Section VI.


\section{Model}
We consider generic hybrid structures where a central unit $H_S$
interacts with two reservoirs $H_{\nu}^0$  ($\nu=L,R$) of
temperatures $T_{\nu}=\beta_{\nu}^{-1}$  ($k_B\equiv 1$) via the coupling terms $V_{\nu}$,
\bea H=H_L^0+H_R^0+H_S+V_L+V_R.
\label{eq:H}
\eea
The heat current from the $\nu$ bath into the subsystem is given by \cite{Current},
$J_{\nu}=\frac{i}{2}{\rm Tr}\left([H_{\nu}^{0}-H_{S},V_{\nu}]\rho \right)$; ($\hbar\equiv 1$), where
$\rho$ is the total density matrix, and we trace over both the subsystem and
reservoirs degrees of freedom. In steady-state the expectation
value of the interaction is zero, ${\rm Tr}\left(\frac{\partial
V_{\nu}}{\partial t}\rho \right)=0$, and we obtain
\bea J_{L}=i{\rm Tr} \left([V_{L},H_{S}]\rho \right); \,\,\
J_{R}=i{\rm Tr} \left([V_{R},H_{S}]\rho \right).
\label{eq:JL1}
\eea
Since in steady-state the currents are equal, $J_L=-J_R$, we
introduce a symmetric definition for the heat current operator
$\widehat{J}$,
\bea J={\rm Tr} [\widehat J \rho]; \,\,\,\,\,\,
\widehat{J}=\frac{i}{2}[V_{L},H_{S}]+\frac{i}{2}[H_{S},V_{R}],
\label{eq:JL} \eea
where the expectation value is defined positive when the current is flowing $L$ to $R$. The
subsystem Hamiltonian assumes a diagonal form,  and we also consider
separable couplings
\bea
H_{S}&=&\sum_n
E_{n}|n\rangle \langle n|;
\nonumber\\
V_{\nu}&=&\lambda_{\nu}SB_{\nu}; \,\,\,\,S=\sum_{n,m}
S_{m,n}|m\rangle \langle n|. 
\label{eq:HSV} 
\eea
Here $S$ is a subsystem operator and $B_{\nu}$ is an operator in
terms of the $\nu$ bath degrees of freedom. For simplicity we set
$S_{m,n}=S_{n,m}$. In what follows we consider situations where
$B_{L}$ and $B_R$ have the same dependence on the (respective) bath
degrees of freedom, with different prefactors $\lambda_{L}\neq
\lambda_{R}$. We refer to this scenario as "parametric asymmetry",
rather than "functional asymmetry", resulting from dissimilar $B$'s.
Note that if the commutator $[H_S,S]=0$,  the heat current trivially
vanishes.

\section{Dynamics}

Employing the Liouville equation in the interaction picture, the
elements of the total density matrix satisfy
\bea \frac{d\rho _{m,n}}{dt}
&=&
-i[V(t),\rho(0)]_{m,n}
\nonumber\\
&-&\int_0^t d\tau
[V(t),[V(\tau),\rho(\tau)]]_{m,n},
\label{eq:Lioville} 
\eea
where $V=V_L+V_R$,  and $V(t)$ are interaction picture operators.
Following standard weak coupling schemes \cite{Breuer}, we proceed by making four assumptions:
(i) We first make the presumption that ${\rm Tr}[V(t),\rho(0)]=0$, i.e. the mean value of
the interaction Hamiltonian, averaged over the initial density matrix, is zero.  
(ii) We factorize the density matrix of the whole system,
at all times, by the product $\rho(t)=\sigma(t) \rho_L(T_L) \rho_R(T_R)$. 
Here $\rho_{\nu}$ is independent of time,
$\rho_{\nu}(T_{\nu})=e^{-H_{\nu}^0/T_{\nu}}/{\rm Tr}_{\nu}[e^{-H_{\nu}^0/T_{\nu}}]$, and
$\sigma(t)$ is the subsystem density matrix obtained by tracing out 
the reservoirs degrees of freedom from the total density matrix, $\sigma(t)={\rm Tr}_{B}[\rho(t)]$.
${\rm Tr}_B$ denotes trace over both the $L$ and the $R$-baths degrees of freedom.
(iii) We take the Markovian limit, 
assuming that the reservoirs' characteristic timescales are 
shorter than the subsystem relaxation time.
(iv) We neglect contributions from quantum coherences, i.e. in the long time limit we assume that
the nondiagonal elements of the system density matrix vanish. This can be justified in the weak 
coupling limit adopting the initial condition $\sigma_{n\neq m}(0)\sim 0$. 
Following these assumptions, the Pauli master equation for the populations 
$P_n(t)={\rm Tr_B}[\rho_{n,n}(t)]$ is obtained \cite{Breuer},
\bea
\dot P_n(t)&=&\sum_{\nu,m}|S_{m,n}|^2P_m(t) k_{m \rightarrow n}^{\nu}(T_{\nu})
\nonumber\\
&-&P_n(t) \sum_{\nu,m} |S_{m,n}|^2 k_{n\rightarrow m}^{\nu}(T_{\nu}).
\label{eq:master}
\eea
The Fermi golden rule transition rates are given by 
\bea
k_{n\rightarrow m}^{\nu}(T_{\nu})=\lambda_{\nu}^2\int_{-\infty }^{\infty
}d\tau e^{iE_{n,m}\tau }\left\langle B_{\nu}(\tau
)B_{\nu}(0)\right\rangle_{T_{\nu}}.
\label{eq:rate0}
\eea
Here  $E_{m,n}=E_m-E_n$,
$\langle B_{\nu}(\tau) B_{\tau}(0)\rangle_{T_{\nu}}=
{\rm Tr}_{\nu} [\rho_{\nu}(T_{\nu}) B_{\nu}(\tau)B_{\tau}(0)]$
is the correlation function of the $\nu$ environment.
In steady-state $\dot P_n=0$, and we normalize the population to
unity $\sum_n P_n=1$. 
It is straightforward to show that under (\ref{eq:HSV}) the steady-state
current (\ref{eq:JL}) becomes
\bea J=\frac{1}{2}\sum_{n,m} E_{m,n}\left\vert S_{m,n}\right\vert
^{2}P_{n}\times [k_{n\rightarrow m}^{L}(T_L) -k_{n\rightarrow m}^{R}(T_R)],
\nonumber\\
\label{eq:current} 
\eea
with the population obtained by solving (\ref{eq:master}) in the long time limit.
For details see Appendix A.
Our description to this point is general, as
we have not yet specified neither the subsystem nor the terminals.

\subsection{Specific models for the subsystem Hamiltonian}

We consider two representative models for the subsystem Hamiltonian
and its interaction with the baths. In the first model the subsystem
is a harmonic oscillator (HO) of frequency $\omega$, $H_S=\sum_n
n\omega|n\rangle \langle n|$. This can describe either a local
radiation mode \cite{Pekola,radiation}, or an active vibrational
mode of the trapped molecule \cite{NDR}. For the interaction
operator we assume $S=\sum_n\sqrt{n} |n\rangle \langle n-1| + c.c$,
motivated by the bilinear form $V_{\nu}\propto xB_{\nu}$,  $x$ is a
subsystem coordinate \cite{NDR}. This implies that only transitions
between nearest states are allowed,
\bea k^{\nu}(T_{\nu})\equiv k_{n\rightarrow n-1}^{\nu}(T_{\nu})&=&\lambda_{\nu}^2\int_{-\infty
}^{\infty }d\tau e^{i\omega \tau }\left\langle B_{\nu}(\tau
)B_{\nu}(0)\right\rangle_{T_{\nu}},
\nonumber\\
k_{n-1\rightarrow n}^{\nu}(T_{\nu})&=&e^{-\beta_{\nu}\omega}k^{\nu}_{n\rightarrow n-1}(T_{\nu}).
\label{eq:raten}
\eea
We also introduce the short notation
\bea k^{\nu}(T_{\nu})=\lambda_{\nu}^2 f_{\nu}(T_{\nu}),
\label{eq:notation}
\eea
where $f_{\nu}$, defined through (\ref{eq:raten}), encompasses the
effect of the bath operators. Solving (\ref{eq:master}) in
steady-state using the above expressions for the rates, the heat
current (\ref{eq:current}) can be analytically calculated \cite{Rectif},
\bea
J^{(HO)}=-\frac{\omega \lbrack n_{B}^{L}(\omega )-n_{B}^{R}(\omega )]}
{n_{B}^{L}(-\omega )/k^{L}(T_L)+n_{B}^{R}(-\omega )/k^{R}(T_R)},
\label{eq:HO}
\eea
where $n_{B}^{\nu}(\omega )=\left[e^{\beta_{\nu}\omega}-1\right]^{-1}$
is the Bose-Einstein distribution function at $T_{\nu}=1/\beta_{\nu}$.

Our second subsystem is a two-level (spin) system (TLS).
Here $H_S=\frac{\omega}{2}\sigma_z$, and we employ a nondiagonal interaction form $S=\sigma_x$.
These terms can represent an electronic spin rotated by the environment
\cite{Lukin}. It can also describe an anharmonic (truncated) molecular vibration that is
dominating  heat flow through the junction \cite{Rectif,NDR}.
Re-calculating the long-time population (\ref{eq:master}),
the heat flux reduces to
\bea
J^{(TLS)}=\frac{\omega \lbrack n_{S}^{L}(\omega )-n_{S}^{R}(\omega )]}{%
n_{S}^{L}(-\omega )/k^{L}(T_L)+n_{S}^{R}(-\omega )/k^{R}(T_R)}
\label{eq:TLS}
\eea
with the rates (\ref{eq:raten}) and the spin occupation factor
$n_{S}^{\nu}(\omega)=\left[e^{\beta_{\nu} \omega}+1\right]^{-1}$.

Expressions (\ref{eq:HO}) and (\ref{eq:TLS}) demonstrate that in the weak
coupling limit the effect of the environment enters only through the
relaxation rates $k^{\nu}$, evaluated at the subsystem energy
spacing $\omega$.
In Section IV we further introduce a three-level system,
an intermediate structure 
between a two-level (strongly anharmonic) system and an 
harmonic object. More generally, given the subsystem anharmonic potential 
and an interaction operator $S$, one should proceed by 
(i) calculating (analytically or numerically) 
the vibrational spectra $|n\rangle$,  (ii) evaluating the matrix elements of $S$, 
(iii) acquiring the steady-state levels populations
by solving (\ref{eq:master}), and (iv) arriving at the heat current using (\ref{eq:current}).

\subsection{Calculation of the rate constant}

We give next the explicit form for the relaxation rate for various
physical thermal baths. The rate constants (\ref{eq:raten}) 
induced by e.g. the $L$ contact is given by the Fourier's transform of the bath
correlation function $\left\langle
B_{L}(\tau)B_{L}(0)\right\rangle_{T_{L}}$, with $B_{L}=\sum
B^{L}_{l,l'}|l\rangle \langle l'|$ and $H_{L}^0=\sum
E_{l}|l\rangle \langle l|$. Here $|l\rangle$ are the many-body
states of the $L$ reservoir with energies $E_l$. 
The rate (\ref{eq:raten}) can be integrated to yield
\bea k^{\nu}(T_{\nu})= 2\pi \lambda_{\nu}^2\sum_{k,k'}\left\vert
B_{k,k'}^{\nu}\right\vert ^{2}\delta (E_{k}-E_{k'}+\omega ) \frac{
e^{-\beta_{\nu} E_k}  } {Z_{\nu}(\beta_{\nu})},
\nonumber\\
\label{eq:ratecalc}
\eea
with $Z_{\nu}(\beta_{\nu})=\sum_k e^{-\beta_{\nu} E_k}$,  the partition
function of the $\nu$ bath; $k,k'\in \nu$.

(i) \textit{Distinguishable noninteracting particles}.
This environment includes a set of independent ($p=1,2,..,P$)
particles. The Hamiltonian is given by summing all separate
contributions
\bea
H_{\nu}^0=\sum_p h^0_{\nu,p}; \,\,\,\, B_{\nu}=\sum_p b_{\nu,p}.
\eea
Within these terms, the relaxation rate (\ref{eq:ratecalc}) reduces to
\bea
k^{\nu}(T_{\nu})&=&
2\pi \lambda_{\nu}^2 \sum_{p}\sum_{i,j}\left\vert \left\langle i\right\vert _{p}b_{\nu,p}\left\vert
j\right\rangle _{p}\right\vert ^{2}
\nonumber\\
&\times&
\delta (\epsilon_{p}(i)-\epsilon_{p}(j)+\omega )
\frac{ e^{-\beta_{\nu} \epsilon_{p}(i)}}{Z_{p}(\beta_{\nu}
)},
\label{eq:R1}
\eea
with the $p$-particle eigenstates $\left\vert i\right\rangle _{p}$
and eigenvalues $\epsilon_{p}(i)$.
$Z_{p}(\beta_{\nu})=\sum_{i}e^{-\beta_{\nu}\epsilon_{p}(i)}$
is the $p$-particle partition function. Specifically, for a bath of noninteracting
spins we get 
\bea
k^{\nu}(T_{\nu})=
\Gamma_S^{\nu}(\omega) n_{S}^{\nu}(-\omega),
\label{eq:R1s}
\eea
with the spin occupation factor 
$n_{S}^{\nu}(\omega)=\left[e^{\beta_{\nu} \omega}+1\right]^{-1}$ and
the temperature independent coefficient
\bea 
\Gamma_S^{\nu}(\omega)=2\pi\lambda_{\nu}^2\sum_{p}
|\left\langle 0\right\vert _{p}b_{\nu,p}\left\vert 1\right\rangle
_{p}| ^{2}\delta (\omega+\epsilon_p(0) -\epsilon_{p}(1)).
\nonumber\\
\eea
%

(ii) \textit{Solid/Radiation field (harmonic bath)}.
This bath includes a set of independent harmonic oscillators, creation operator
$a_{\nu,j}^{\dagger}$. System-bath interactions are further assumed to be bilinear,
\bea H_{\nu}^0=\sum_{j}\omega_j a_{\nu,j}^{\dagger} a_{\nu,j};
\,\,\,\,\, B_{\nu}=\sum_{j}
(a_{\nu,j}+a_{\nu,j}^{\dagger }).
\label{eq:HR2} 
\eea
This leads to the relaxation rate 
\bea
k^{\nu}(T_{\nu})=-\Gamma_B^{\nu}(\omega)n_{B}^{\nu}(-\omega),
\label{eq:R2}
\eea
with the Bose-Einstein function $n_B^{\nu}(\omega)=[e^{\beta_{\nu}\omega}-1]^{-1}$ and
an effective system-bath coupling factor
\bea \Gamma_B^{\nu}(\omega)&=&2\pi \lambda_{\nu}^2\sum_j \delta
(\omega _{j}-\omega).
 \eea
%
(iii) \textit{Fermionic bath: metal}.
Consider a  metallic terminal including a set of noninteracting spinless electrons, 
creation operator $c_{\nu,i}^{\dagger}$,
where only scattering between electronic states within the same lead are allowed,
\bea
H_{\nu}^0=\sum_i \epsilon _{i}c_{\nu,i}^{\dagger }c_{\nu,i}; \,\,\,\
B_{\nu}=\sum_{i,j} c_{\nu,i}^{\dagger }c_{\nu,j}.
\eea
The transition rate between subsystem states (\ref{eq:ratecalc}) can be written as \cite{Persson}
\bea
k^{\nu}(T_{\nu}) &=&-2\pi\lambda_{\nu}^2 n_{B}^{\nu}(-\omega ) \sum_{i,j} \delta (\epsilon
_{i}-\epsilon _{j}+\omega )
\nonumber\\
&&\times \lbrack n_{F}^{\nu}(\epsilon_{i})-n_{F}^{\nu}(\epsilon _{i}+\omega )],
\eea
with the Fermi-Dirac distribution function
$n_F^{\nu}(\epsilon)=[e^{\beta_{\nu}(\epsilon-\mu_{\nu})}+1]^{-1}$,
$\mu_{\nu}$ is the chemical potential. One could also write
\bea
k^{\nu}(T_{\nu})=- \Lambda^{\nu}(T_{\nu},\omega)
n_{B}^{\nu}(-\omega),
\label{eq:R4}
\eea
where
\bea
\Lambda^{\nu}(T_{\nu,}\omega)=
2\pi \int d\epsilon \left[ n_{F}^{\nu}(\epsilon )-n_{F}^{\nu}(\epsilon +\omega )\right]F_{\nu}(\epsilon).
\label{eq:FGR}
\eea
The function $F_{\nu}(\epsilon)=\lambda_{\nu}^2\sum \delta (\epsilon
-\epsilon _{j}+\omega )\delta (\epsilon_{i}-\epsilon)$
depends on the system-bath coupling elements and the specific band structure.
Assuming that the density of states slowly varies in the (subsystem) energy
window $\omega$, the interaction function can be expanded around the
chemical potential  \cite{Persson}, $F_{\nu}(\epsilon)\approx
F_{\nu}(\mu_{\nu})+ \gamma_{\nu}
\frac{|\epsilon|-\mu_{\nu}}{\mu_{\nu}}$, with $\gamma_{\nu}$ a
dimensionless number of order unity. Using this form, the
integration in Eq. (\ref{eq:FGR}) can be performed when the Fermi
energy is much bigger than the conduction band edge $\mu_{\nu}\gg
E_c$. One can then write
\bea
\Lambda^{\nu}(T_{\nu,}\omega)\approx \Gamma_F^{\nu}(\omega) \left(1+\delta_{\nu} \frac{T_{\nu}}{\mu_{\nu} }\right),
\label{eq:R4M}
\eea
where  $\delta$ is a constant of order one, measuring the deviation from a
flat band structure near the chemical potential $\mu_{\nu}$ \cite{radiation}.
The coupling constant is given by
\beq 
\Gamma_{F}^{\nu}(\omega)= 2\pi\omega F_{\nu}(\mu_{\nu}). 
\eeq 
When $\delta_{\nu}=0$ we find that $\Lambda ^{\nu}(T_{\nu},\omega)= \Gamma_F^{\nu}(\omega)$, a temperature independent
constant, and  the harmonic limit (\ref{eq:R2}) is retrieved \cite{radiation}.

To conclude this discussion, assuming different types of reservoirs and
system-bath couplings, we derived here three expressions for the
relaxation rate (\ref{eq:ratecalc}) [or equivalently (\ref{eq:raten})]: 
Eq.  (\ref{eq:R1s}) assuming a spin bath,  Eq. (\ref{eq:R2}) for a  phononic
environment with a linear coupling to the subsystem, and
Eq. (\ref{eq:R4}) for a metallic bulk with electron-hole pair excitations
\beq
k^{\nu}(T_{\nu})=
\begin{cases}
n^{\nu}_S(-\omega)\Gamma_S^{\nu}(\omega); \,\,\,\,\,\,\,\, \rm {noninteracting \,\,spins} \,\\
-n^{\nu}_B(-\omega)\Gamma_B^{\nu}(\omega); \,\, \rm{ phonons; \, linear \, coupling}\, \\
-n_B^{\nu}(-\omega)
(1+\delta_{\nu}\frac{T_{\nu}}{\mu_{\nu}}) \Gamma_F^{\nu}(\omega)
 \,\,\,\,\,\, \rm{Metal}\,
\end{cases}
\label{eq:summ}
\eeq
In each case the coefficient  $\Gamma^{\nu}$  reflects
the system-bath interaction strength, whereas the temperature dependent
function encloses the reservoirs properties.


\section{Nonlinear current-temperature characteristics}
Based on the two models for the subsystem (HO, TLS)  and
the different types of baths, we can construct several two-terminal
heat-conducting junctions. We consider next few examples, and manifest 
nonlinear current-temperature behavior. 
We will also introduce a three-level system (3LS), an intermediate structure
between a two-level (strongly anharmonic) system and an harmonic object. 
Formally, if we write the heat current as 
\bea
J(T_a,\Delta T)=\sum_{k}
\alpha_k(T_a)\Delta T^k
\label{eq:formal}
\eea
with $\Delta T=T_L-T_R$; $T_a=T_L+T_R$, we ask
ourselves what are the microscopic parameters that are incorporated
in the nonlinear coefficients $\alpha_k$; $k>1$, and how can we
control the magnitude of these terms.
We also define the thermal conductance as $\mathcal K=\lim_{\Delta T \to 0}J/\Delta T$,
and examine its temperature dependence. 

1. {\it Harmonic Junction}. Our first example is a fully harmonic
system, incorporating two harmonic reservoirs connected by a harmonic link, 
modeling vibrational/photonic heat transfer between two solids/ohmic metals 
\cite{SegalHanggi,Pekola,Ojanen},
assuming that a specific harmonic mode of frequency $\omega$
dominates the dynamics. The current is given by (\ref{eq:HO}) with
the rates (\ref{eq:R2}), resulting in the thermal Landauer
expression \cite{Kirczenow},
\begin{subequations} \label{eq:R2HOR2}
\begin {align}
&J=  \mathcal T_B \omega
\left[n_B^L(\omega)-n_B^R(\omega)\right] \label{eq:R2HOR2a}\\
& \stackrel {\beta_{\nu}\omega \ll 1}  {\longrightarrow}
\mathcal T_B
\left[ \Delta T -\frac{\omega^2}{3T_a^4}\Delta T^3 + O(\Delta
T^5)\right]. \label{eq:R2HOR2b}
\end{align}
\end{subequations}
Here $\mathcal T_B=\frac{\Gamma_B^L \Gamma_B^R}{\Gamma_B^L+\Gamma_B^R}$ 
is a (temperature independent)
transmission coefficient, with the coupling elements
$\Gamma_B^{\nu}$ calculated at the molecular frequency $\omega$.
This equation describes ballistic thermal transport, where energy
loss takes place only at the contacts. Note that in the classical
limit $J \propto \Delta T$, $\mathcal K = \mathcal T_B$, 
and the current does not directly depend
on the molecular (subsystem) vibrational frequency. For a harmonic
junction nonlinear effects are thus purely quantum, originating from
the quantum statistics.

2. {\it Spin Junction}. Our second example is a spin-TLS-spin system
representing e.g. a central electron interacting with two
nuclear-spin environments under an applied magnetic field
\cite{Lukin}. The current is given by Eq. (\ref{eq:TLS}) with the
rates (\ref{eq:R1s}), bringing in a spin-Landauer formula,
\begin{subequations} \label{eq:R1TLSR1}
\begin{align}
&J=  \mathcal T_S \omega  \left[
n_S^L(\omega)-n_S^R(\omega)\right] \label{eq:R1TLSR1a}\\
& \stackrel {\omega\beta_{\nu} \ll 1}  {\longrightarrow} \mathcal T_S
\frac{\omega^2}{T_a}\sum_{n=1,3,5...}\left(\frac{\Delta
T}{T_a}\right)^{n}, \label{eq:R1TLSR1b}
\end{align}
\end{subequations}
with the transmission coefficient $\mathcal T_S=\frac{\Gamma_S^L
\Gamma_S^R}{\Gamma_S^L+\Gamma_S^R}$, $\Gamma_S^{\nu}$ is evaluated
at $\omega$, the central spin energy spacing. Unlike
(\ref{eq:R2HOR2b}), the linear response term here does depend on the
subsystem frequency, $\mathcal K= \mathcal T_S \omega^2/T_a^2$, and nonlinear terms
persist even in the high temperature limit.
Note that both Eqs. (\ref{eq:R2HOR2a}) and (\ref{eq:R1TLSR1a}) 
are in the form of Landauer formula, 
since only elastic scattering processes are effective (system and reservoirs are identical).
Our next two examples bring in deviations from the Landauer picture.

3. {\it Harmonic baths-spin subsystem junction}. Consider again
vibrational heat flow. However, unlike the fully harmonic model 
resulting in (\ref{eq:R2HOR2}), the central unit is assumed here
to incorporate anharmonic terms \cite{Rectif, radiation, Tomi}. This can be implemented by modeling
the subsystem by a truncated harmonic oscillator, for example a spin. The current across the
device is evaluated using (\ref{eq:TLS}) with the rates (\ref{eq:R2}),
\begin{subequations}  \label{eq:R2TLSR2}
\begin{align}
&J=\omega
\frac{\Gamma_B^L\Gamma_B^R(n_B^L(\omega)-n_B^R(\omega))}{\Gamma_B^L(1+2n_B^L(\omega))+\Gamma_B^R(1+2n_B^R(\omega))} \label{eq:R2TLSR2a}\\
& \stackrel {\omega\beta_{\nu}\ll1}  {\longrightarrow} \mathcal T_B
\omega \sum_{n=1}^{\infty} \left(\frac{\Delta T}{T_a}\right)^n
(-\chi_B)^{n-1}, \label{eq:R2TLSR2b}
\end{align}
\end{subequations}
where $\chi_B=\frac{\Gamma_B^L-\Gamma_B^R}{\Gamma_B^L+\Gamma_B^R}$
measures the spatial asymmetry.
Note that we could still organize Eq. (\ref{eq:R2TLSR2a}) in the
form of the Landauer formula with a {\it temperature dependent} transmission
coefficient \cite{Tu}. The high-temperature linear response limit yields
$\mathcal K = \mathcal T_B\omega/T_a $.
Nonlinear terms survive in (\ref{eq:R2TLSR2b}) {\it only due to the asymmetric coupling},
since for $\chi_B=0$, $J\propto \Delta T$.
Thus, quite interestingly, this anharmonic junction conducts linearly (in the classical limit),
as long as it is fully symmetric.

\begin{figure}[htbp]
\hspace{2mm} {\hbox{\epsfxsize=90mm \epsffile{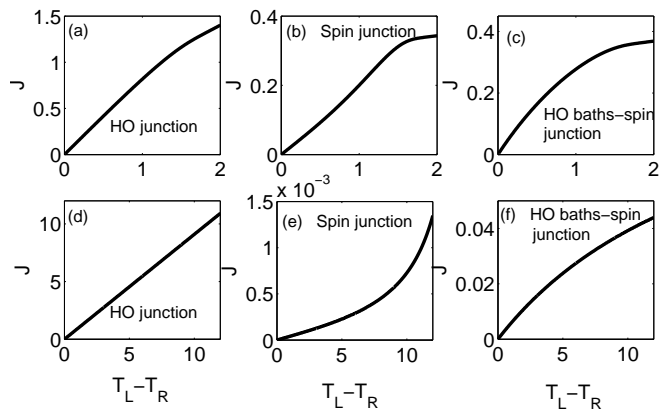}}} \caption{
Nonlinear heat flow in various hybrid junctions. (a)-(c):
$\omega=1$, $T_a=2$.  (d)-(f): $\omega=0.1$, $T_a=15$. (a) and (d)
are harmonic junctions with the current (\ref{eq:R2HOR2}); (b) and
(e) are spin junctions obeying (\ref{eq:R1TLSR1}); (c) and (f) are
harmonic baths-spin junctions obeying (\ref{eq:R2TLSR2}).
$\Gamma^{L}$=10, $\Gamma^{R}=1$ in all figures. The
$x$ axis is $\Delta T$. } \label{non1}
\end{figure}

In Fig. \ref{non1} we exemplify the properties of the junctions
(\ref{eq:R2HOR2})-(\ref{eq:R2TLSR2}) 
with asymmetric couplings $\Gamma^L\neq\Gamma^R$, taking
$\omega$, the subsystem characteristic energy, to be either of the
order of the reservoirs temperature, $\omega \sim T_a$, or
significantly lower. In the first case subplots (a)-(c) reveal that
the current saturates at large $\Delta T$, reflecting the quantum
statistics. In contrast, in the high temperature limit, while a pure harmonic
system (d) shows a linear current-$\Delta T$ characteristic, the other
two systems (e)-(f), incorporating some nonlinearities,  reveal
nonlinear behavior.

%

4. {\it Three-level junctions}.
In order to further elucidate the role of a uniform energy spectra
in nonlinear transport,
we construct next a three-level system (3LS), an intermediate structure
between a two-level (strongly anharmonic) subsystem and an 
harmonic object. 
We assume that both the subsystem and the reservoirs include (equivalent) 3LS particles,
$H_S=\sum_{n=1,2,3}E_n|n\rangle \langle n|$;
$S=|1\rangle\langle 2|+|2\rangle\langle 3| +h.c.$.
The current  (\ref{eq:current}) reduces to
\bea
J&=&\frac{\omega_1}{D_1} k_{2\rightarrow 1}^L  k_{2\rightarrow 1}^R(e^{-\beta_L\omega_1}-e^{-\beta_R\omega_1}) 
\nonumber\\
&+& \frac{\omega_2}{D_2} k_{3\rightarrow 2}^L  k_{3\rightarrow 2}^R(e^{-\beta_L\omega_2}-e^{-\beta_R\omega_2}),
\eea
with  $\omega_1=E_2-E_1$; $\omega_2=E_3-E_2$, and
\bea
D_1&=&k_{1\rightarrow 2} k_{2\rightarrow 3}/k_{3\rightarrow 2}+k_{1\rightarrow 2}+k_{2\rightarrow 1};
\nonumber\\
D_2&=&k_{3\rightarrow 2} k_{2\rightarrow 1}/k_{1\rightarrow 2}+k_{3\rightarrow 2}+k_{2\rightarrow 3}.
\eea
Here $k_{i\rightarrow j} = k_{i\rightarrow j}^L +k_{i\rightarrow j}^R$, with the rates
defined in (\ref{eq:rate0}).
Taking $\omega=\omega_1=\omega_2$, and assuming the reservoirs include collections of 3LS noninteracting particles 
of equal spacing $\omega$, we obtain the following expression, using (\ref{eq:R1}) ($\omega\ll T_a$),
\bea
J= \frac{16}{9}\frac{\omega^2}{T_a}\mathcal T_S \sum_{n=1,3,5...} \left(\frac{\Delta T}{T_a}\right) ^n
\label{eq:J3LS}
\eea
with $\mathcal T_S=\frac{\Gamma_S^L\Gamma_S^R}{\Gamma_S^L+\Gamma_S^R}$ and
$\Gamma_S^{\nu}=2\pi \lambda_{\nu}^2 \sum_p|\langle i|b_{\nu,p}|j\rangle_p|^2 \delta(\omega+\epsilon_p(i)-
\epsilon_p(j))$; $i<j$.
This expression is essentially similar to (\ref{eq:R1TLSR1}).
It is notable that only {\it odd} terms survive, irrespective of the spatial asymmetry ($\Gamma^L\neq \Gamma^R$). It can be shown
that even terms participate only when the energy spectra along the device becomes inhomogeneous.

5. {\it Energy flow between metals}.
Energy transfer between metals, mediated by the excitation of a single radiation mode,
has recently attracted considerable experimental and theoretical interest
\cite{Pekola,radiation,Tomi,Ojanen}.
Analogous junctions are the core of  molecular electronics,  where the bridging modes are the vibrations
of the trapped molecules.
For example, heat dissipation in a metal surface-C$_{60}$-STM junction
was demonstrated to be controlled by vibrational decay into the tip, generating electron-hole pair excitations \cite{C60,Persson}. 
Heat conduction through a DNA-gold composite, suspended between two electrodes, was found to be dominated by phonon transport \cite{DNA}.

When both metals are ohmic, the
current in the weak coupling limit is given by (\ref{eq:R2HOR2}) \cite{radiation}.
As we show next, for structured reservoirs nonlinear effects emerge.
We calculate the heat current using Eq. (\ref{eq:HO}) with the rates (\ref{eq:R4})-(\ref{eq:R4M}),
assuming $\mu=\mu_{\nu}$ and $\delta=\delta_{\nu}$, i.e. the reservoirs are equivalent,
and are maintained at the same chemical potential. This leads to
\begin{subequations} \label{eq:R4HOR4}
\begin{align}
&J= \omega \frac{\Gamma_F^R\Gamma_F^L(1+\delta\frac{T_L}{\mu})
(1+\delta\frac{T_R}{\mu}) \left[n_B^L(\omega)-n_B^R(\omega)\right] }
{\Gamma_F^L(1+\delta\frac{T_L}{\mu})+
\Gamma_F^R(1+\delta\frac{T_R}{\mu})}  \label{eq:R4HOR4a}\\
&\stackrel {\omega\beta_{\nu} \ll 1}{\longrightarrow} \mathcal T_F
\Big[ \left(1+\delta \frac{T_a}{2\mu}\right)\Delta T - \chi_F
\frac{\delta}{2\mu} \Delta T^2 \nonumber\\ &-
\frac{\delta^2}{\mu^2}\frac{\Gamma_F^L\Gamma_F^R}{(\Gamma_F^L+\Gamma_F^R)^2}
\Delta T^3+ O(\Delta T^4) \Big]. \label{eq:R4HOR4b}
\end{align}
\end{subequations}
In deriving (\ref{eq:R4HOR4b}) we assumed that $\delta T_a/2\mu <1$.
Here $\chi_F=\frac{\Gamma_F^L-\Gamma_F^R}{\Gamma_F^L+\Gamma_F^R}$
measures the asymmetry in the system-bath coupling and $\mathcal
T_F=\frac{\Gamma_F^L\Gamma_F^R}{\Gamma_F^L+\Gamma_F^R}$. The
following observations can be made: (i) The nonlinear contributions are all related
to a finite $\delta$, measuring the deviation from a constant
density of states \cite{radiation}. If $\delta=0$, the harmonic
limit is recovered. (ii)  The existence of even terms, e.g. a $\Delta
T^2$ term, requires some asymmetry, $\chi_F\neq0$, as we discuss below,
see Eq. (\ref{eq:rectifR4HOR4}). (iii) The conductance of the junction scales like
$\mathcal K= \mathcal T_{F}(1+\delta T_a/2\mu)$, in sharp contrast to the behavior of
phononic systems. 

\begin{figure}[htbp]
\hspace{2mm} {\hbox{\epsfxsize=85mm \epsffile{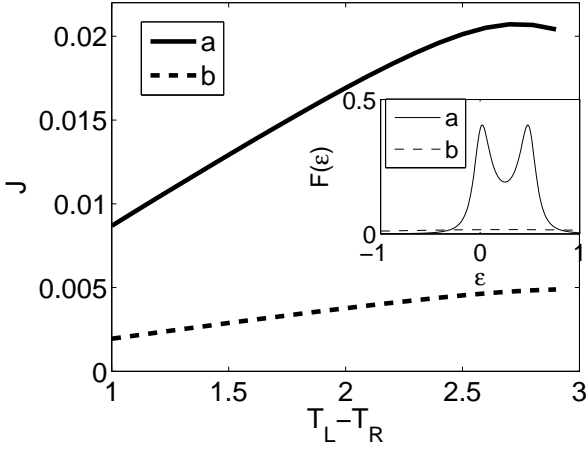}}} \caption{
Negative differential conductance in a metal-single mode- metal
junction with $\gamma=0.2$ (solid line). $\Gamma_F^{L}$=$\Gamma_F^{R}$=1,
$T_a=3$, $\omega=0.5$. The dashed curve was generated using
$\gamma=5$, manifesting behavior similar to  the fully harmonic case (\ref{eq:R2HOR2}).
Inset: The function $F(\epsilon)$ [see (\ref{eq:FGR})] vs. energy
for $\gamma=0.2$ (solid line); $\gamma=5$ (dashed line).} \label{non2}
\end{figure}

In Fig. \ref{non2} we study energy transfer between metals employing
a Lorentzian density of states of width $\gamma$,
$D_{\nu}(\epsilon)=\frac{\gamma/2\pi}{\epsilon^2+\gamma/4^2}$. We
observe a small 'negative differential conductance', a decrease of the
current with increasing temperature different at large temperature
bias. The effect prevails when $\gamma\lesssim \omega$, i.e. the
reservoirs density of states is notably changing over the
subsystem energy scale. The data was generated by employing
(\ref{eq:HO}) with the rate (\ref{eq:R4}). The inset presents the
$F$ function (\ref{eq:FGR}), evaluated with the Lorentzian DOS.

Summarizing, while in pure harmonic systems nonlinear 
dynamics is linked to quantum effects (\ref{eq:R2HOR2}), 
spin systems manifest nonlinear dynamics 
irrespective of the subsystem frequency, (\ref{eq:R1TLSR1}) and (\ref{eq:J3LS}). 
In a simple model for anharmonic vibrational heat flow 
(\ref{eq:R2TLSR2}) nonlinearities are attributed to the spatial asymmetry,
while nonlinear effects in excitonic energy transfer are linked to 
the metals' energy dependent density of states (\ref{eq:R4HOR4}).
The conductance temperature dependence also varies. For harmonic systems it
is independent of temperature, while for anharmonic phononic systems it decays like $1/T$,
in general agreement with experimental results  \cite{Pop}.
In metal-single mode-metal junctions the conductance {\it increases} with temperature, 
due to the enhancement of the transition rate with $T$.

%


\section{Thermal rectification: Sufficient conditions}

We focus next on a specific nonlinear effect, thermal
rectification, asymmetry of the heat current for forward and 
reversed temperature difference,  $|J(+\Delta T)|\neq |J(-\Delta T)|$. 
Formally, the onset of this phenomenon is identified by the existence of even terms,
$\alpha_{k=2n}\neq 0$; $n=1,2...$, in the expansion (\ref{eq:formal}).

We discuss here sufficient conditions for thermal rectification
by analyzing the currents in (\ref{eq:HO}) and (\ref{eq:TLS}). 
These expressions are naturally not the most general results for heat transfer
across hybrid structures. Rather, they relay on few assumptions as explained
in Section III: The subsystem and reservoirs
are weakly interacting, the baths are assumed to be held at thermal equilibrium, and
the dynamics is markovian.
Furthermore, specific forms for the subsystem are employed. 
However, these analytical forms, derived from a prototype model, can still 
illuminate on the basic mechanisms involved in thermal rectification.
Having said so, we return to Eqs. (\ref{eq:HO}) and (\ref{eq:TLS}), and analyze their
structure for forward ($T_L=T_H$; $T_R=T_C$) and reversed ($T_L=T_C$; $T_R=T_H$) temperature
gradients, $T_H>T_C$.
We note that in each expression separately these currents 
deviate in magnitude if the denominators fulfill
\bea \frac{n^{H}(-\omega) }{k^L(T_H)} +
\frac{n^{C}(-\omega)}{k^R(T_C)} \neq \frac{n^{C}(-\omega)
}{k^L(T_C)} + \frac{n^{H}(-\omega)}{k^R(T_H)}, \eea
where $n^{\nu}$ is either the spin occupation factor or the
Bose-Einstein distribution. Rearranging this expression and making use of
 (\ref{eq:notation}) we get
\bea \frac{n^{H}(-\omega) }{\lambda_{L}^2f_L(T_H)} -
\frac{n^{H}(-\omega) }{\lambda^2_{R}f_R(T_H)}  \neq
\frac{n^{C}(-\omega) }{\lambda_{L}^2f_L(T_C)} -
\frac{n^{C}(-\omega)}{\lambda_{R}^2f_R(T_C)}.
\label{eq:condd} \eea
In what follows we identify two classes of rectifiers. In type-A
rectifiers system-bath interactions are equivalent at both contacts,
but the reservoirs have distinct properties. In type-B rectifiers
the reservoirs are equivalent, but the subsystem statistics is
distinct from the baths, combined with some parametric asymmetry.

\subsection{Type-A thermal rectifier}

The rate constants (\ref{eq:ratecalc}) can be expressed in terms of the
reservoirs density of states. For example, the rate induced by
the $L$ contact is given by 
\bea
k^{L}(T)&=&
2\pi \lambda_L^2\sum_{l,l'}\left\vert B_{l,l'}^L\right\vert ^{2}\delta (E_{l}-E_{l'}+\omega )
\frac{e^{-\beta E_{l}}}{Z_L(\beta )}
\nonumber\\
&=& 2\pi \lambda_L^2 \frac{\int d\epsilon e^{-\beta
\epsilon}D_L(\epsilon) g_L(\epsilon,\omega)} {\int d\epsilon
e^{-\beta \epsilon}D_L(\epsilon)}. \label{eq:rateG} 
\eea
Here $Z_L(\beta)=\sum_l e^{-\beta E_l}=\int d\epsilon e^{-\beta
\epsilon}D_L(\epsilon)$ denotes the partition function of the $L$
bath with the density of states
$D_L(\epsilon)=\sum_l\delta(\epsilon-E_l)$. The function
$g_L(\epsilon,\omega)=\sum_{l}|B^L(\epsilon,E_l)|^2\delta(\epsilon-E_l+\omega)$
characterizes system-bath interactions.
Taking $\lambda_L=\lambda_R$, we note that since the two sides of
(\ref{eq:condd}) depend on different temperatures, the system
rectifies heat if
\bea 
f_L(T)\neq f_R(T), 
\label{eq:ineq} 
\eea
besides special points in the parameter space, depending on the
details of the model. Using (\ref{eq:rateG}), this condition reduces
to
\bea
\frac{\int d\epsilon e^{-\beta \epsilon} D_L(\epsilon) g_L(\epsilon,\omega) }
{\int d\epsilon e^{-\beta \epsilon}D_L(\epsilon)}
\neq \frac{\int d\epsilon e^{-\beta \epsilon} D_R(\epsilon) g_R(\epsilon,\omega) }
{\int d\epsilon e^{-\beta \epsilon}D_R(\epsilon)}.
\nonumber\\
\label{eq:typeA}
\eea
We next further assume that
$g_L(\epsilon,\omega)=g_R(\epsilon,\omega)$, i.e. system-bath
interaction matrix elements are equal at both ends. Hence, the inequality
(\ref{eq:ineq}) is satisfied when
\beq
D_{L}(\epsilon)\neq D_R(\epsilon).
\eeq
We thus recover a sufficient condition for thermal rectification:
The density of states of the reservoirs should be distinct. Note
that at least one of the reservoirs should have an {\it energy
dependent} DOS. If both reservoirs are harmonic, $D_{\nu}=c_{\nu}$,
with $c_{\nu}$ a constant, rectification is absent even for $c_L\neq
c_R$. It is thus sufficient to have one of the reservoirs
incorporating anharmonic interactions. If both reservoirs are
nonlinear, rectification takes place if the energy spectra are
distinct. This discussion relays on the assumption that the spectral
function $g(\epsilon, \omega)$ explicitly depends on energy, see Eq.
(\ref{eq:typeA}) and \cite{comment}.

Going back to the condition (\ref{eq:typeA}), we 
Taylor-expand the interaction function around
$\omega$, $g_{\nu}(\epsilon,\omega)\sim
\alpha_1(\omega)+\alpha_2(\omega)\epsilon$,
\bea
\frac{\int d\epsilon e^{-\beta \epsilon} D_L(\epsilon) \epsilon}
{\int d\epsilon e^{-\beta \epsilon}D_L(\epsilon)}
\neq \frac{\int d\epsilon e^{-\beta \epsilon} D_R(\epsilon) \epsilon }
{\int d\epsilon e^{-\beta \epsilon}D_R(\epsilon)}.
\eea
We identify the left (right) hand side as the average energy of the
$L$ ($R$) reservoir. This relation manifests  that rectification
exists if $\langle H_L^0\rangle \neq \langle H_R^0\rangle$, as
in \cite{Sufficient}. Specifically, consider interfaces including 
1-dimensional oscillator chains,
$H_{\nu}^0=H_{\nu}^{kin}+H_{\nu}^{pot}$, where the kinetic energy
$H_{\nu}^{kin}$ is quadratic in momentum, and the potential energy
per particle is $C_{n}q^n$; $n\geq 2$, $q$ is the particle's
coordinate. In the classical limit using the equipartition relation
we obtain $\langle H_{\nu}^0 \rangle =
T_{\nu}(\frac{1}{2}+\frac{1}{n})$. Thermal rectification thus
emerges if the reservoirs have a non-identical power $n$.
%
Note that the separation to three segments ($L$, subsystem, $R$) is
often artificial: The junction could be practically made of a single
structure with a varying potential energy, e.g. a nanotube whose
composition gradually changes in space \cite{RectifE}, leading to
inhomogeneous energy spectra, thus to thermal rectification.

To conclude, rectification has been obtained here relaying on the
DOS asymmetry $D_L(\epsilon) \neq D_R(\epsilon)$, while system-bath
interactions are assumed symmetric, $\lambda_L=\lambda_R$ and
$g_L(\omega,\epsilon)=g_R(\omega,\epsilon)$,  energy dependent
functions. This conclusion is in accord with multitude (numerical) observations,
demonstrating rectification in two-segment dissimilar anharmonic lattices
\cite{Casati,Zhang,Rectif2D,Rectif3D,RectifMass}.


\subsection{Type-B thermal rectifier}
We explore next the role of the subsystem statistics in inducing
thermal rectification by further studying the inequality
(\ref{eq:condd}). As discussed above, type-A rectifiers are
constructed assuming that $f_L(T)\neq f_R(T)$, resulting from the
use of dissimilar reservoirs. However, a more careful analysis of
(\ref{eq:condd}) reveals that rectification prevails even when
$f(T)\equiv f_L(T)=f_R(T)$, as long as $\lambda_{L}\neq \lambda_R$
{\it and} the ratio $n^{\nu}(-\omega)/f_{L,R}(T_{\nu})$ depends on
the temperature $T_{\nu}$. We show it by rearranging
(\ref{eq:condd}),
\bea \frac{n^{H}(-\omega) }{f(T_H)} \left( \frac{1}{\lambda_L^2} -
\frac{1}{\lambda_R^2}\right) \neq \frac{n^{C}(-\omega) }{f(T_C)}
\left( \frac{1}{\lambda_L^2} - \frac{1}{\lambda_R^2}\right). \eea
In order for the two sides to deviate, the ratio, e.g.,
$n^{H}(-\omega)/f(T_{H})$  must depend on the respective temperature.
In other words, {\it the relaxation rates' temperature
dependence should differ from the central unit particle statistics}.
As shown in Section III.B, the function $f_{\nu}(T)$ reflects the reservoirs
statistics, see Eqs. (\ref{eq:notation}) and  (\ref{eq:summ}).
We therefore classify type-B rectifiers as junctions
where subsystem and bath differ in their statistics, and the
identical reservoirs are asymmetrically coupled to the subsystem
$\lambda_L\neq \lambda_R$. 
This observation is in agreement with other studies.
For example, in Ref. \cite{radiation} rectification was demonstrated  
in a quantum-mechanical model where photon-mediated heat current 
flows between two (asymmetrically coupled) nonlinear reservoirs.
Refs. \cite{Rectif} and \cite{Tomi}  consider
a nonlinear subsystem mode while the reservoirs are taken harmonic, 
yielding rectification due to the inclusion of 
some spatial asymmetry.

Summarizing, in type-B rectifiers the reservoirs and systems-bath
couplings are equal at both ends, $D_L(\epsilon)=D_R(\epsilon)$ and
$g_L(\epsilon,\omega)=g_R(\epsilon, \omega)$, but (i) one of the
contacts is weaker, $\lambda_L\neq \lambda_R$, and (ii) the
subsystem statistics differs from the reservoirs' statistics.


\subsection{Examples}

In what follows we exemplify type-A and type-B rectifiers.
The magnitude of thermal rectification can be estimated by the sum $\Delta J\equiv
J_{\Delta T}+J_{-\Delta T}$, where $J_{\pm \Delta T}=J(T_L-T_R=\pm
\Delta T)$, or by the ratio $\mathcal R\equiv |J_{\Delta T}
/J_{-\Delta T}|$.
As expected from our general analysis above, a fully
harmonic system (\ref{eq:R2HOR2}), a pure spin system
(\ref{eq:R1TLSR1})  and a uniform 3LS model  (\ref{eq:J3LS}) 
do not rectify heat {\it irrespective of
the spatial asymmetry} embodied in $\chi\neq0$, since (i) the DOS of both contacts are equal,
and (ii) subsystem and baths are equivalent.
This should be emphasized since the latter two cases can be viewed as anharmonic structures, due to the truncated energy spectra. 
In contrast, for a harmonic bath- TLS-harmonic bath junction (\ref{eq:R2TLSR2}),  we obtain the ratio
\bea
\mathcal R\sim 1-2 \frac{\Delta T}{T_a}\chi_B.
\eea
Similarly, for a metal- single mode-metal junction, using
(\ref{eq:R4HOR4}) at small $\delta$, we find
\bea
\mathcal R\sim 1 -\frac{\delta}{\mu} \chi_F \Delta T.
\label{eq:rectifR4HOR4}
\eea
Thus, in type-B rectifiers the strength of the effect is directly
linked to the spatial asymmetry, $\chi\neq 0$ \cite{Rectif, radiation}.

We demonstrate next some type-A rectifiers, where the reservoirs have distinct properties. 
In order to simplify the presentation we set
$\Gamma=\Gamma_B^{L}=\Gamma_S^R$, i.e. the central unit is 
evenly coupled to the reservoirs. Our first  example is a phonon bath- HO-
spin bath junction, representing e.g. an insulating molecule
interfacing  a dielectric surface and a nuclear-spin environment.
For a schematical representation see Fig. \ref{Fig0}(a).  The
current is calculated using (\ref{eq:HO}) with the rates
(\ref{eq:R1s}) and (\ref{eq:R2}) resulting in
\bea
J=\Gamma \omega \frac{n_B^L(\omega)-n_B^R(\omega)}{1 -
\frac{n_B^R(-\omega)}{n_S^R(-\omega)}}.
\eea
The magnitude of thermal rectification is
\bea 
\mathcal{R} =\frac{ n_{B}^{H}(-\omega )}{n_{B}^{C}(-\omega
)}=\frac{1-\exp (-\frac{2\omega }{ T_a-\Delta T })}{1-\exp(-\frac{2\omega }{T_a+\Delta T })}.
\eea
It is easy to see that when $\frac{2\omega}{T_a\pm \Delta T}\gg1$,
the rectifier is effectively turned off while for $\frac{2\omega
}{T_a\pm \Delta T}\ll 1$ it is operative, with $\mathcal{R}\sim
1+2\Delta T/T_a$.

Our second example is a phonon bath-HO-metal junction,
representing an electronic to vibrational energy conversion device,
see Fig. \ref{Fig0}(b). This system can be realized in a metal-molecule contact,
where the vibrational decay into the lead (e.g. an STM tip) 
is bottlenecked by a single vibrational mode.
Setting the coupling strength at both
contacts to be the same, $\Gamma=\Gamma_B^{L}=\Gamma_F^R$, we get
\bea
J= \omega \Gamma \frac{n_B^L(\omega)-n_B^R(\omega)}{1 + \left(1+\delta_R \frac{T_R}{\mu_R}\right)^{-1}},
\eea
with the rectification ratio
\bea
\mathcal R=
\frac{(2+\delta_R \frac{T_H}{\mu_R})(1 +\delta_R \frac{T_C}{\mu_R})}
{(1 +\delta_{R}\frac{ T_H}{\mu_R})(2 +\delta_R \frac{T_C}{\mu_R})}
\sim 1-\delta_R\frac{\Delta T}{2\mu_R}.
\eea
This expression is similar to (\ref{eq:rectifR4HOR4}).
However, since here the reservoirs are distinct (effectively $\delta_L=0$, $\delta_R\neq0$),
rectification is obtained for $\chi=0$.

\begin{figure}[htbp]
\hspace{2mm}
{\hbox{\epsfxsize=90mm \epsffile{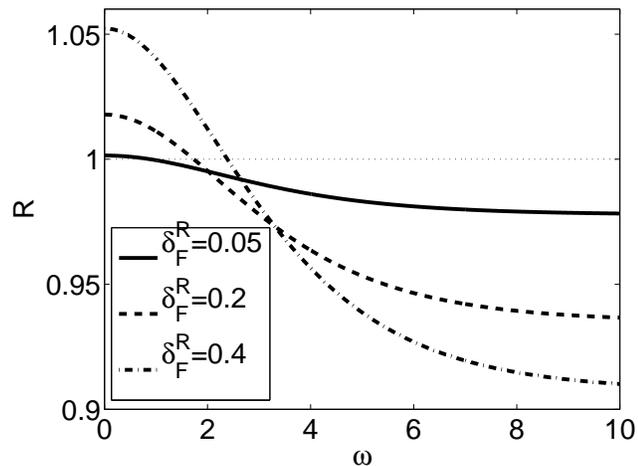}}}
\caption{Electronic to vibrational energy exchange through a TLS mode
with $\delta_R=0.05$ (solid line); $\delta_R=0.2$ (dashed line); $\delta_R=0.4$ (dashed-dotted line).
Rectification ratio is presented as a function of the subsystem energy spacing.
$T_a=3$, $\Delta T=1$, $\mu_{R}=1$, $\Gamma_B^L=\Gamma_F^R=1$.
}
\label{Fig2}
\end{figure}

Fig. \ref{Fig2} presents a similar instance,  vibrations to
electronic excitations energy exchange, mediated by the excitation of an {\it anharmonic molecular} (TLS)
mode. Interestingly, for small $\omega$ the solid transfers heat to the metal more
effectively that the reversed metal-to-dielectric process, reflected by a
rectification ratio larger than 1. For large spacing the behavior is
reversed, and the metal better cools down. Consequently, there is an
intermediate value ($\omega\sim3$ for $\delta_F^R=0.4$) where this
nonlinear-inhomogeneous junction does not rectify heat.
The rectification ratio obtained here is relatively small,
since the metal only weakly deviates from the ohmic description \cite{radiation}.
This example still demonstrates that by optimizing system and interface parameters
it may be possible to design a junction where the electron bath can be
efficiently cooled down, while the vibrational energy 
ineffectively dissipates into the metallic bulk, and vice-versa.

\begin{figure}[htbp]
\hspace{2mm}
{\hbox{\epsfxsize=90mm \epsffile{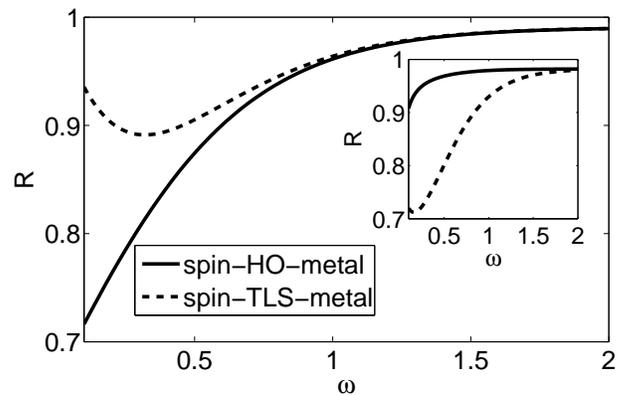}}}
\caption{Spin-HO-metal rectifier (solid line) and a
spin-TLS-metal rectifier (dashed line). The rectification ratio is
presented as a function of the subsystem energy spacing. $T_a=0.5$,
$\Delta T=0.1$, $\delta_{R}=0.2$, $\mu_{R}=1$, Main plot: The
subsystem is equally coupled to the two ends,
$\Gamma_S^L=\Gamma_F^R=1$. Inset: $\mathcal R$ can be tuned by
manipulating system-bath interactions, $\Gamma_S^L=1$,
$\Gamma_F^R=0.05$. } \label{Fig1}
\end{figure}

Finally, we examine a spin-subsystem-metal junction. 
In Fig. \ref{Fig1} we show that it can rectify heat in the classical
limit ($\omega<T_{\nu}$) while in the quantum regime rectification is
suppressed. We also modify the system-metal coupling strength, and
show that it can largely control the rectification ratio (inset).
%


\section {Summary}
This paper provided a unified description of heat flow in two-terminal hybrid 
structures assuming weak system-bath couplings.
The underlying origin of nonlinear behavior in various junctions, and its controllability,
were explored considering different interfaces: metals, insulators and
noninteracting spins, where the central object (subsystem) could
represent e.g., a radiation mode or a vibrational excitation.

As a particular example of nonlinear conduction we examined thermal rectification.
Previous studies of this effect were typically based on numerics,
utilizing classical molecular dynamics tools, 
where observations were deduced within  specific molecular force fields
 \cite{Terraneo,Casati,Zhang, Rectif2D, Rectif3D, RectifMass,Zeng,Prosen}.
Here, in contrast, we attempted a general analytical study of
the sufficient conditions for the onset of thermal rectification in 
hybrid quantum models. We identified two classes of
rectifiers: Type-A rectifiers where the interfaces are made
distinct, and type-B rectifiers where the reservoirs are equivalent,
but the system and bath quantum statistics differ, in conjunction
with some spatial (parametric) asymmetry.
We note that the importance of anharmonicity and asymmetry 
for manifesting rectification  were previously recognized, e.g. in a
spin-boson model \cite{Rectif}, yet the necessity of the
inhomogeneity of the energy spectra was not appreciated. 
Here we clearly observe that a system composed of identical {\it anharmonic} 
units cannot rectify heat, unless the energy spectra are made {\it non-uniform}, 
in conjunction with some spatial asymmetry.

Our study aims in linking transport characteristics to the microscopic Hamiltonian.
It might serve as a guide for experimentalists pursuing control over  
energy transfer in molecular systems \cite{C60,DNA,Hamm,Dlott},
and for building nanoscale thermal devices \cite{Baowenlogic,Tunable}.
We have also demonstrated that nonlinear thermal effects, and in particular thermal rectification,
are ubiquitous phenomena that could be observed in a variety of systems, phononic \cite{RectifE},
electronic \cite{QDrectif}, and photonic \cite{Pekola,radiation,Ojanen}.

\begin{acknowledgments}
This work was supported by
the University of Toronto Start-up Funds.
L.-A. Wu acknowledges support  from the Ikerbasque foundation.
\end{acknowledgments}


\renewcommand{\theequation}{A\arabic{equation}}
\setcounter{equation}{0}  
\section*{APPENDIX A: Derivation of the  heat current expression within the master equation formalism}
The  aim of this Appendix is to derive the heat current expression (\ref{eq:current}) 
within the Hamiltonian (\ref{eq:HSV}).
The expectation value of the current is formally given by
\bea J={\rm Tr} [\widehat J \rho]; \,\,\,\,\,\,
\widehat{J}=\frac{i}{2}[V_{L},H_{S}]+\frac{i}{2}[H_{S},V_{R}].
\label{eq:AJL} 
\eea
Using the  Hamiltonian (\ref{eq:HSV}) we get
\bea
J&=&\frac{i}{2}\sum_{n,m} E_{n,m}S_{m,n} {\rm Tr}_B[\lambda_L\rho_{n,m}B_L]  
\nonumber\\
&-&\frac{i}{2}\sum_{n,m} E_{n,m}S_{m,n} {\rm Tr}_B[\lambda_R\rho_{n,m}B_R],
\label{eq:Ac1}
\eea
where $E_{n,m}=E_n-E_m$, ${\rm Tr}_B$ denotes trace over the $L$ and $R$ baths, 
and $\rho$ is the total density matrix whose
 elements should be calculated in the long time limit,
since  expression  (\ref{eq:AJL})  is valid in steady-state situations only \cite{Current}.
We find it useful to rearrange Eq.  (\ref{eq:Ac1}) as follows,
\bea
J&=&\frac{i}{2}\sum_{n>m} E_{n,m}S_{m,n} {\rm Tr_B}\big[\lambda_L(\rho_{n,m}-\rho_{m,n})B_L\big]
\nonumber\\
&-&\frac{i}{2}\sum_{n>m} E_{n,m}S_{m,n} {\rm Tr_B}\big[\lambda_R(\rho_{n,m}-\rho_{m,n})B_R\big].
\nonumber\\
\label{eq:Ac2}
\eea
We would like to express the current in terms of the long-time population $P_n$.
The Liouville equation of motion for $\rho_{n,m}$ is 
\bea
&&\dot \rho_{n,m}=
-iE_{n,m}\rho_{n,m} 
\nonumber\\
&&-i\sum_{\nu}\sum_p \lambda_{\nu} \big[B_{\nu}(t) S_{n,p}\rho_{p,m}(t) - S_{p,m}\rho_{n,p}(t) B_{\nu}(t)\big],
\nonumber\\
\eea
with the transformed operators $B_{\nu}(t)=e^{iH_{\nu}t}B_{\nu}e^{-iH_{\nu}t}$;
$\nu=L,R$.
The index $p$ counts  the subsystem states.
This equation can be formally integrated to yield
\bea
\rho_{n,m}(t)&=&-i\int_0^t e^{-iE_{n,m}(t-\tau)} \Big [
\sum_{\nu,p}\lambda_{\nu} B_{\nu}(\tau) S_{n,p}\rho_{p,m}(\tau)
\nonumber\\
&-&\sum_{\nu,p} \lambda_{\nu} S_{p,m}\rho_{n,p}(\tau) B_{\nu}(\tau)  \Big ]d\tau.
\label{eq:Ac3}
\eea
The initial condition was already neglected since it will not contribute after tracing out 
the bath in (\ref{eq:Ac2}), assuming  ${\rm Tr}_B[B(t),\rho(0)]=0$, i.e. the mean value of
the interaction Hamiltonian, averaged over the initial density matrix, is zero.
Plugging  (\ref{eq:Ac3}) into  (\ref{eq:Ac2}) we obtain $\lambda^2$ order terms.
Factorizing the density matrix at all times,  $\rho(t)=\rho_L(T_L) \rho_R(T_R) \sigma(t)$,
$\rho_{\nu}(T_{\nu})=e^{-H_{\nu}^0/T_{\nu}}/{\rm Tr}_{\nu}[e^{-H_{\nu}^0/T_{\nu}}]$; $\sigma(t)={\rm Tr}_{B}[\rho(t)]$,
and taking the markovian limit, replacing $\sigma(\tau)\rightarrow
\sigma(t)$, we get
\bea
&&{\rm Tr}_{B}[\rho_{n,m}(t)B_L(t)]\approx
\nonumber\\
&&-i\lambda_L \sum_p\int_0^t e^{-iE_{n,m}(t-\tau)} \Big [ \langle B_L(t) B_L(\tau) \rangle_{T_L}
\sigma_{p,m}(t) S_{n,p}  
\nonumber\\
&&- \langle B_L(\tau)B_L(t)\rangle_{T_L} \sigma_{n,p}(t)S_{p,m} \Big] d\tau.
\label{eq:Ac4}
\eea
Here $\langle B_{\nu}(\tau) B_{\tau}(0)\rangle_{T_{\nu}}\equiv
{\rm Tr}_{\nu} [\rho_{\nu}(T_{\nu}) B_{\nu}(\tau)B_{\tau}(0)]$
is the correlation function of the $\nu$ environment.
In deriving (\ref{eq:Ac4}) we disregarded mixed terms of the form  $\langle B_L(t)B_R(\tau)\rangle$,
since the terminals are not correlated.
Our next assumption is that coherences are negligible in the weak-coupling scheme, given
the preparation $\sigma_{n\neq m}(0)\sim 0$, thus
we disregard all the nondiagonal terms in  (\ref{eq:Ac4}). 
Further, performing the transformation $x=\tau-t$ and extending the
integral to infinity (markovian approximation), we obtain
\bea
&&{\rm Tr}_{B}[\rho_{n,m}(t)B_L(t)]\approx
\nonumber\\
&&-i\lambda_L \int_{-\infty}^0 e^{iE_{n,m}x} \Big [ \langle B_L(0) B_L(x) \rangle_{T_L}
\sigma_{m,m}(t) S_{n,m}  
\nonumber\\
&&- \langle B_L(x)B_L(0)\rangle_{T_L} \sigma_{n,n}(t)S_{n,m} \Big] dx.
\label{eq:Ac5}
\eea
Similarly, we find that
\bea
&&{\rm Tr}_{B}[\rho_{m,n}(t)B_L(t)]\approx
\nonumber\\
&&-i\lambda_L \int_0^{\infty} e^{iE_{n,m}x} \Bigg [ \langle B_L(x) B_L(0) \rangle_{T_L}
\sigma_{n,n}(t) S_{m,n}  
\nonumber\\
&&- \langle B_L(0)B_L(x)\rangle_{T_L} \sigma_{m,m}(t)S_{m,n} \Bigg] dx.
\label{eq:Ac5b}
\eea
Combining the last two expressions, we come by (using $S_{n,m}=S_{m,n}$)
\bea
&&{\rm Tr}_{B}[(\rho_{n,m}(t)-\rho_{m,n}(t))B_L(t)]\approx
\nonumber\\
&& i\lambda_L S_{n,m} \Bigg[ P_n(t)\int_{-\infty}^{\infty} e^{iE_{n,m}x} \langle B_L(x) B_L(0)\rangle_{T_L}
\nonumber\\
&& - P_m(t)\int_{-\infty}^{\infty} e^{iE_{n,m}x} \langle B_L(0) B_L(x)\rangle_{T_L}\Bigg].
\label{eq:Ac5c}
\eea
We return now to the heat current expression (\ref{eq:Ac2}), 
and identify the $\nu$-bath induced relaxation rates by
\bea
k_{n\rightarrow m}^{\nu}(T_{\nu})=\lambda_{\nu}^2\int_{-\infty }^{\infty
}d\tau e^{iE_{n,m}\tau }\left\langle B_{\nu}(\tau
)B_{\nu}(0)\right\rangle_{T_{\nu}}.
\eea
Making use of (\ref{eq:Ac5c}), conjoined with the analogous $R$-bath expression,
we get a weak-coupling expression for the heat current, defined positive when
flowing $L$ to $R$,
\bea
J&=&-\frac{1}{2}\sum_{n>m}E_{n,m} |S_{m,n}|^2 \big[P_n k_{n\rightarrow m}^L(T_L)-P_m k_{m\rightarrow n}^L(T_L) \big]
\nonumber\\
&+&
\frac{1}{2}\sum_{n>m}E_{n,m} |S_{m,n}|^2 \big[P_n k_{n\rightarrow m}^R(T_R)-P_m k_{m\rightarrow n}^R (T_R) \big],
\nonumber\\
\eea
or more compactly
\bea
J=\frac{1}{2}\sum_{n,m}E_{m,n}|S_{n,m}|^2 P_n  \big[k_{n\rightarrow m}^L(T_L) - k_{n\rightarrow m}^R(T_R)\big].
\nonumber\\
\label{eq:Ac6}
\eea
This is our final result, a second-order ($\lambda^2$) expression for the heat current, obtained 
within the master equation approach.
The populations $P_n$ should be calculated at long time, using (\ref{eq:master}).



\begin{thebibliography}{99}
\bibitem{comment}
The function $g_{\nu}(\epsilon,\omega)$ typically depends on energy. For example,
assuming a harmonic bath and bilinear interactions,  (\ref{eq:HR2}),
we get $g(\epsilon, \omega)=\sum_l l \delta(\epsilon-E_{l-1}) \delta(E_{l-1}-E_l+\omega)$.
\end{thebibliography}
\end{document}